\RequirePackage{ifpdf}
\ifpdf 
\documentclass[pdftex]{sigma}
\else
\documentclass{sigma}
\fi

\def\IC{\mathbb{C}}

\def\IZ{{\mathbb{Z}}}
\def\IR{{\mathbb{R}}}
\def\IP{\mathbb{P}}

\def\CL {{\cal L}}

\def\CH {{\cal H}}
\def\one{{\hbox{ 1\kern-.8mm l}}}

\def\pdbar{\dot{\bar{\phi}} }
\def\jbar{\bar{\jmath}}

\begin{document}

\allowdisplaybreaks

\renewcommand{\thefootnote}{$\star$}

\renewcommand{\PaperNumber}{069}

\FirstPageHeading

\ShortArticleName{Balanced Metrics and Noncommutative K\"ahler Geometry}

\ArticleName{Balanced Metrics\\ and Noncommutative K\"ahler Geometry\footnote{This paper is a
contribution to the Special Issue ``Noncommutative Spaces and Fields''. The
full collection is available at
\href{http://www.emis.de/journals/SIGMA/noncommutative.html}{http://www.emis.de/journals/SIGMA/noncommutative.html}}}

\Author{Sergio LUKI\'C}

\AuthorNameForHeading{S.~Luki\'c}

\Address{Department of Physics and Astronomy, Rutgers University, Piscataway, NJ 08855-0849, USA}
\Email{\href{mailto:lukic@physics.rutgers.edu}{lukic@physics.rutgers.edu}}

\ArticleDates{Received March 01, 2010, in f\/inal form August 02, 2010;  Published online August 27, 2010}

\Abstract{In this paper we show how Einstein metrics are naturally described using the quantization of the algebra of functions $C^{\infty}(M)$ on a K\"ahler manifold $M$. In this setup one interprets $M$ as the phase space itself, equipped with the Poisson brackets inherited from the K\"ahler 2-form. We compare the geometric
quantization framework with several deformation quantization approaches. We f\/ind that the \emph{balanced metrics}
appear naturally as a result of
requiring the vacuum energy to be the constant function on the moduli space of \emph{semiclassical vacua}. In the classical limit these metrics become K\"ahler--Einstein (when~$M$ admits such metrics). Finally, we sketch several applications of this formalism, such as explicit constructions of special Lagrangian submanifolds in compact Calabi--Yau manifolds.}


\Classification{14J32; 32Q15; 32Q20; 53C25; 53D50} 

\renewcommand{\thefootnote}{\arabic{footnote}}
\setcounter{footnote}{0}

\vspace{-2mm}

\section{Introduction}

Noncommutative deformations of K\"ahler geometry exhibit some extraordinary features, similar to those that appear in the description of $n$ quantum harmonic oscillators by the noncommutative phase space $\IC^n$.
Noncommutative geometry in Calabi--Yau compactif\/ications is expected to play a special role when
the $B$-f\/ield is turned on \cite{Seiberg:1999vs, Kapustin:2003sg}, in the formulation of M(atrix) theory
\cite{Banks:1996vh, Cornalba:1998zy, Kachru:1997bi}, and in the large $N$ limit of probe D0-branes
\cite{Gaiotto:2004pc}. Also, as we show below, one can use the geometric quantization approach to noncommutative geometry to determine\footnote{More precisely, what one can determine are certain metrics, known as balanced metrics, which obey the equations of motion in the classical limit.} important objects in string theory compactif\/ications, which allow the computation of the exact form of the Lagrangian in the four dimensional ef\/fective f\/ield theory \cite{Don:i, Don:iii, DKLR:i, DKLR:ii}.

In this paper we show how the notion of \emph{balanced metrics} appears naturally in the framework of K\"ahler quantization/noncommutative geometry. In the geometric quantization formalism the balanced metric appears as a consequence of requiring the norm of the coherent states to be constant; these states are parameterized by the Kodaira's embedding of the K\"ahler manifold $M$ into the projectivized quantum Hilbert space. In the Reshetikhin--Takhtajan approach to the deformation quantization of $M$ \cite{nikolai}, the balanced metric appears as a consequence of requiring the unit of the quantized algebra of functions $C^{\infty}(M)[[\hbar]]$ to be the constant function ${\bf 1}\colon M\to 1\in\IR$. Finally, we quantize the phase space with constant classical Hamiltonian on $M$ using the path integral formalism; here, one considers a dif\/ferent class of semiclassical vacuum states which are not coherent states, and uses them to def\/ine\footnote{By requiring the vacuum energy density to be the constant function on the semiclassical vacua, labeled by the points $x \in M$.} a generalization of the balanced metrics (which also become Einstein metrics in the \emph{classical limit}).

The organization of the paper is as follows. In Section~\ref{section2} we recall some basic facts of geometric quantization \cite{Axelrod:1989xt, Elitzur:1989, Souriau:1986}, show how the balanced metrics appear naturally in this framework, and sketch how dif\/ferential geometric objects, such as K\"ahler--Einstein metrics or Lagrangian submanifolds, can be described in this language. In Section~\ref{section3} we summarize the work of Reshetikhin--Takhtajan, and show how the constant function is the unit element of their quantized algebra of functions if and only if the metric on $M$ is balanced. Finally, in Section~\ref{section4} we consider a dif\/ferent set of semiclassical vacuum states in path integral quantization and use them to def\/ine \emph{generalized balanced metrics}, which dif\/fer slightly from the balanced metrics in geometric quantization.

\section{Geometric quantization}\label{section2}

Classical mechanics and geometric quantization have a beautiful formulation using the language of symplectic geometry, vector bundles, and operator algebras \cite{Axelrod:1989xt, Elitzur:1989, Souriau:1986}. In this language, symplectic manifolds $M$ are interpreted as phase spaces, and spaces of smooth functions $C^{\infty}(M)$ as the corresponding classical observables.

K\"ahler quantization is understood far better than quantization on general symplectic manifolds; for this reason we only consider K\"ahler manifolds (which are symplectic manifolds endowed with a compatible complex structure). $(M, \CL^{\otimes \kappa})$ denotes a polarized K\"ahler manifold $M$ with a very ample hermitian line bundle\footnote{In other words, $\CL^{\otimes\kappa}$ is an element of the K\"ahler cone associated to $M$.} $\CL^{\otimes \kappa}$, and $\kappa\in \IZ_+$ a positive integer. For technical reasons, we consider $M$ to be compact and simply connected. We work with a trivialization of $\CL\vert_U\to U$, where $U\subset M$ is an open subset; we def\/ine $K(\phi,\bar{\phi})$ to be the associated analytic K\"ahler potential and $e^{-\kappa K(\phi,\bar{\phi})}$ the hermitian metric on $\CL^{\otimes \kappa} \to M$. If ${\rm dim}_{\IC}\, M=n$ and $\{\phi_i \}_{0< i \leq n}$ is a local holomorphic coordinate chart for the open subset $U\subset M$, we can write the K\"ahler metric on~$M$ and its compatible symplectic form as
\begin{gather*}
i \kappa g_{i\bar{\jmath}} d\phi^i \otimes d\bar{\phi}^{\bar{\jmath}} = \kappa \omega_{i\bar{\jmath}} d\phi^i \wedge d\bar{\phi}^{\bar{\jmath}} = i\kappa {\partial\over \partial \phi^i} {\partial\over \partial \bar{\phi}^{\bar{\jmath}}}
K(\phi,\bar{\phi}) d\phi^i \otimes d\bar{\phi}^{\bar{\jmath}}.
\end{gather*}

Classically, the space $(C^{\infty}(M),  \omega)$ of observables has, in addition to a Lie algebra structure def\/ined by the Poisson bracket
\[
\{f,  g \}_{PB}= \omega^{i\bar{\jmath}}(\partial_i f \bar{\partial}_{\bar{\jmath}} g-\partial_i g\bar{\partial}_{\bar{\jmath}} f), \qquad f,  g \in C^{\infty}(M),
\]
the structure of a commutative algebra under pointwise multiplication,
\[
(fg)(x) = f(x)g(x) = (gf)(x).
\]
Quantization can be understood as a non-commutative deformation of $C^{\infty}(M)$ parameterized by $\hbar$, with commutativity recovered when $\hbar = 0$. We will discuss the formalism of deformation quantization in the next section, although generally speaking, quantization refers to an assignment $T\colon f\to T(f)$ of classical observables to operators on some Hilbert space $\CH$. When $M$ is compact, the Hilbert space will be f\/inite-dimensional with dimension $
\dim \CH = {{\rm vol}\,M \over \hbar^n}+O(\hbar^{1-n})$. The assignment $T$ must satisfy the following requirements:
\begin{itemize}\itemsep=0pt
\item Linearity, $T(af + g) = aT(f) + T(g)$, $\forall \, a\in \IC$, $f,g\in C^{\infty}(M)$.
\item Constant map $1$ is mapped to the identity operator ${\rm Id}$, $T(1)={\rm Id}$.
\item If $f$ is a real function, $T(f)$ is a hermitian operator.
\item In the limit $\hbar \to 0$, the Poisson algebra is recovered $[T(f),T(g)]=i\hbar T(\{f,g\}_{PB}) + O(\hbar^2)$.
\end{itemize}

In geometric quantization the positive line bundle $\CL^{\otimes\kappa}$ is known as \emph{prequantum line bundle}. The prequantum line bundle is endowed with a unitary connection whose curvature is the
symplectic form $\kappa\omega$ (which is quantized, i.e., $\omega\in H^{2}(M,\IZ)$). The \emph{prequantum Hilbert space} is the space of~$L^2$ sections
\[
L^2(\CL^{\otimes\kappa}, M) = \left\{ s\in \Omega^{0}(\CL^{\otimes\kappa})\colon \  \int_{M} h^\kappa \langle s,\bar{s}\rangle {\omega^{n}\over n!}<\infty \right\},
\]
where $h^\kappa$ is the compatible hermitian metric on $\CL^{\otimes\kappa}$. The Hilbert space is merely a subspace of~$L^2(\CL^{\otimes\kappa}, M)$, def\/ined with the choice of a polarization on $M$. In the case of K\"ahler polarization, the split of the tangent space in holomorphic and anti-holomorphic directions, $TM = TM^{(1,0)}\oplus TM^{(0,1)}$, def\/ines a Dolbeault operator on $\CL^{\otimes\kappa}$, $\bar{\partial}\colon \Omega^{(0)}(\CL^{\otimes\kappa})\to \Omega^{(0,1)}(\CL^{\otimes\kappa})$. The Hilbert space~$\CH_\kappa$ is only the kernel of $\bar{\partial}$, i.e., the space of holomorphic sections $H^{0}(M, \CL^{\otimes\kappa})$.

As a f\/inal remark, the quantization map $T$ is not uniquely def\/ined; there are dif\/ferent assignments of smooth functions on~$M$ to matrices on~$\CH_{\kappa}$ that obey the same requirements stated above, giving rise to equivalent classical limits. For simplicity, we mention only the most standard ones~\cite{Bordemann:i}:
\begin{itemize}\itemsep=0pt
\item \emph{The Toeplitz map:}
$ T(f)_{\alpha\bar{\beta}} = \int_{M} f(z,\bar{z}) s_{\alpha}(z)
\bar{s}_{\bar{\beta}}(\bar{z}) h^{\kappa}(z,\bar{z}) {\omega(z,\bar{z})^{n}\over n!}, $ with $s_{\alpha}$ a basis of sections for $\CH_{\kappa}$ and $s_{\alpha}(z)$ the corresponding evaluation of $s_\alpha$ at $z\in U\subset M$.
\item \emph{The geometric quantization map:} $Q(f) = iT\left(f - {1\over 2}\Delta f\right) $, with $\Delta$ the corresponding Laplacian on $M$.
\end{itemize}
We will work only with completely degenerated Hamiltonian systems (i.e.\ a constant Hamiltonian function on $M$); therefore the choice of quantization map will not be important. Rather we will study the semiclassical limit of the corresponding quantized system by determining the semiclassical vacuum states.

\subsection{Coherent states and balanced metrics}

As we described above, the geometric quantization picture is characterized by the prequantum line bundle,
$\CL^{\otimes \kappa}\to M$, a holomorphic line bundle on $M$ which is endowed with a $U(1)$ connection with K\"ahler 2-forms $\kappa\omega$. As the positive integer $\kappa$ always appears multiplying the symplectic form, one can interpret $\kappa^{-1}=\hbar$ as a discretized Planck's constant. Thus, according to this convention, the semiclassical appears in the limit $\kappa\to\infty$.

In the local trivialization $U\subset M$, where $K(\phi,\bar{\phi})$ is the K\"ahler potential and $e^{-\kappa K(\phi,\bar{\phi})}$ the hermitian metric on $\CL^{\otimes\kappa}\vert_U$, one can set the compatible Dolbeault operator to be locally trivial and write the covariant derivative as
\[
\widetilde{\nabla} = d\phi^i(\partial_{i} - \kappa\partial_i K) + d\bar{\phi}^{\bar{\imath}}\bar{\partial}_{\bar{\imath}},
\]
where $K$ is the yet undetermined analytic K\"ahler potential on $\CL$.
One can also determine the associated unitary connection up to a $U(1)$ gauge transformation,
\[
\nabla = d\phi^i(\partial_{i} + A_i) + d\bar{\phi}^{\bar{\imath}}(\bar{\partial}_{\bar{\imath}}-A_{i}^\dagger),
\]
with $A_i = \sqrt{h^{-\kappa}}\partial_i \sqrt{h^\kappa}$, and $h=\exp(-K(\phi,\bar{\phi}))$.

As explained above, the Hilbert space $\CH_\kappa$ corresponds to the kernel of the covariant half-derivative $\nabla^{(0,1)}\colon \Omega^{(0)}(\CL)\to \Omega^{(0,1)}(\CL)$,
which are the holomorphic sections of $\CL^{\otimes \kappa}$
\[
\CH_\kappa = H^0(M, \CL^{\otimes \kappa}) = {\rm span}_{\IC}\left\{\vert s_\alpha \rangle \right\}_{\alpha=1}^N.
\]
The dimension of the quantum Hilbert space is
\[
N = \dim \CH_\kappa = {1\over n!}\int_M c_1(\CL)^n \kappa^n + {1\over 2(n-1)!}
\int_M c_1(\CL)^{n-1}c_1(M)\kappa^{n-1} + O\big(\kappa^{n-2}\big).
\]

We identify $\vert s_\alpha\rangle$ as the basis elements of $\CH_\kappa$.
The \emph{coherent state} localized at $x\in M$ can be def\/ined (see~\cite{Rawnsley:1976}) on the trivialization $\CL^{\otimes\kappa}\vert_U\to U\subset M$ as the ray in $\IP\CH_\kappa$ generated by
\[
\vert \tilde\Omega_x \rangle = \sum_\alpha s_\alpha(x)\exp (-\kappa K(x,\bar{x})/2) \vert s_\alpha\rangle \in \CH_\kappa,
\]
where $s_\alpha(x)\exp (-\kappa K(x,\bar{x})/2)$ is the evaluation of the holomorphic section
$\vert s_\alpha\rangle$ at the point $x\in U\subset M$, in the trivialization $\CL^{\otimes\kappa}\vert_U$.
The coherent states are a supercomplete basis of $\CH_\kappa$, and obey the Parseval identity
\begin{gather}
\label{parseval}
\langle \zeta \vert \xi \rangle = \int_{M\hookrightarrow \IP\CH_\kappa} \langle \zeta \vert \tilde \Omega_x \rangle  \langle \tilde \Omega_x \vert \xi \rangle {\omega^n(x,\bar{x}) \over n!},\qquad \forall\, \zeta, \xi\in\CH_\kappa.
\end{gather}
These points in $\IP\CH_\kappa$ are independent of the trivialization,  and they have the property of being
localized at $x\in M$ with minimal quantum uncertainty.
The \emph{distortion function}, diagonal of the Bergman kernel, or expected value of the identity at $x$, $\rho(x,\bar{x})$ is def\/ined as
\begin{gather}
\label{distortion}
\rho = \langle \tilde\Omega_x \vert \tilde\Omega_x \rangle = \sum_{\alpha, \beta} \bar{s}_{\bar{\alpha}}(\bar{x}) s_{\beta}(x) \exp \left( -\kappa K(x,\bar{x})\right) \langle s_\alpha \vert s_\beta \rangle,
\end{gather}
which measures the relative normalization of the coherent states located at dif\/ferent points of $M$. Imposing $\rho(x,\bar{x})= \langle \tilde\Omega_x \vert \tilde\Omega_x \rangle =\mathrm{const}$,
constrains the K\"ahler potential $K(x,\bar{x})$ to be a~Fubini--Study K\"ahler potential:
\begin{gather}
\label{FS}
K(x,\bar{x}) = {1\over \kappa}\log \left( \sum_{\alpha, \beta} \bar{s}_{\bar{\alpha}}(\bar{x}) s_{\beta}(x) \langle s_\alpha \vert s_\beta \rangle \right).
\end{gather}
One of the most important ingredients in the quantization procedure is the def\/inition of the
quantization map, $T:C^{\infty}(M)\to {\rm Herm}(\CH_\kappa)$. This maps classical observables,
i.e.\ smooth real functions on the phase space $X$, to quantum
observables, i.e., self-adjoint operators on the Hilbert space $\CH_\kappa$.
If we work with an orthonormal basis $\langle s_\beta \vert s_\alpha\rangle = \delta_{\beta\alpha}$,
the quantization condition
\[
T(1_{M}) = \mathrm{Id} \in \CH_\kappa\otimes\CH_\kappa^\ast
\]
implies that the embedding of the coherent states satisf\/ies the \emph{balanced} condition \cite{Don:iii},
\begin{gather}
\label{norm}
\delta_{\alpha\beta}=\langle s_\alpha \vert s_\beta \rangle = \sum_x \langle s_\alpha \vert \tilde\Omega_x \rangle \langle \tilde\Omega_x \vert s_\beta\rangle = \int_M {\bar{s}_{\alpha}(\bar{x})s_\beta(x) \over \sum_\gamma \vert s_\gamma(x)\vert^2}   {\omega(x,\bar{x})^n\over n!};
\end{gather}
here, we have used the Parseval identity \eqref{parseval}, and the Liouville's volume form on the phase space~$M$, which can be written as
\[
{1\over n!} \omega(z,\bar{z})^n = {1\over n!} \left[  \bar{\partial}_{\bar{\jmath}} \partial_i K(z,\bar{z}) dz^{i}\wedge d\bar{z}^{\bar{\jmath}}\right]^n.
\]

In summary, in the geometric quantization of an algebraic K\"ahler manifold, the homogeneity of the distortion function $\langle \tilde\Omega_x \vert \tilde\Omega_x \rangle$ and the mapping of the constant function on $M$ to the identity operator ${\rm Id}\colon \CH_\kappa\to \CH_\kappa$, determines a unique metric on $M$ known as \emph{balanced metric}. In the semiclassical limit, $\kappa\to\infty$, this sequence of balanced metrics approaches the K\"ahler--Einstein metric (if it exists) as sketched below (see~\cite{Don:i, DKLR:ii}).

\subsection{Emergence of classical geometry}

For every $\kappa$, the \emph{balanced metric} has just been def\/ined as result of
requiring $\langle \tilde\Omega_x \vert \tilde\Omega_x \rangle$ to be the constant function on $M$.
In the semiclassical limit, $\kappa \to\infty$, we can expand the distortion function in inverse powers of $\kappa$ (see \cite{Zelditch:1998})
\[
\langle \tilde\Omega_{x,\kappa} \vert \tilde\Omega_{x,\kappa} \rangle \sim 1 + {1\over 2\kappa}R + O\big(\kappa^{-2}\big) + \cdots,
\]
and therefore the sequence of balanced metrics will converge to a metric of constant scalar curvature at $\kappa=\infty$. For a Calabi--Yau manifold this is equivalent to a Ricci f\/lat K\"ahler metric.
It is interesting to note that if the identity matrix is identif\/ied with the quantum Hamiltonian, and the coherent states with the semiclassical states, the \emph{balanced metric} can also be def\/ined as the metric that yields a constant semiclassical vacuum energy $\langle \tilde\Omega_{x,\kappa} \vert \tilde\Omega_{x,\kappa} \rangle$, as a function of $x\in M$ and f\/ixed $\kappa$.

Other geometrical elements that one can recover naturally are the Lagrangian submanifolds with respect the K\"ahler--Einstein symplectic form. In the K\"ahler $n$-fold $(M,\omega)$, the level sets of~$n$
commuting functions $(f_1, f_2,\ldots, f_n)$ under the Poisson bracket
\[
\{f_a,f_b \}_{PB}= \omega^{i\bar{\jmath}}(\partial_i f_a\bar{\partial}_{\bar{\jmath}} f_b-\partial_i f_b\bar{\partial}_{\bar{\jmath}} f_a)=0, \qquad \forall\, a,b,
\]
def\/ine a foliation by Lagrangian submanifolds. One can recover such commutation relations as the
classical limit of $n$ commuting self-adjoint operators on the Hilbert space $\CH_\kappa$~\cite{Bordemann:i}:
\[
\langle \tilde\Omega_{x,\kappa}\vert [\hat{f}_a,\hat{f}_b] \vert \tilde\Omega_{x,\kappa} \rangle \sim {i \over \kappa}\{f_a,f_b \}_{PB} + O(\kappa^{-2}),
\]
with $\langle \tilde\Omega_{x,\kappa} \vert \hat{f}_a \vert \tilde\Omega_{x,\kappa} \rangle \to f_a(x)$, and $\vert \tilde\Omega_{x,\kappa} \rangle$ the coherent state peaked at $x\in X$. Thus, one can approximate Lagrangian
submanifolds by using $n$-tuples of commuting matrices for large enough $\kappa$. One can impose further conditions, i.e.\ $\mathrm{Im}(\Omega)\vert_{\rm SLag}=0$, in order to describe special Lagrangian submanifolds. More precisely,
we def\/ine the quantum operator
\begin{gather*}
{\cal I}_{\bar{\alpha}\beta \bar{\alpha}_1\beta_1\dots \bar{\alpha}_n\beta_n} = {1\over 2i} \int_{M} {\omega^n \over n!}
\bar{s}_{\bar{\alpha}}s_{\beta} {\rm e}^{-\kappa K}
\Big( \Omega_{i_1 \dots i_n}\partial^{i_1}\big(\bar{s}_{\bar{\alpha}_1}s_{\beta_1} {\rm e}^{-\kappa K}\big)\cdots\partial^{i_n}\big(\bar{s}_{\bar{\alpha}_n}s_{\beta_n} {\rm e}^{-\kappa K}\big) \nonumber \\
\phantom{{\cal I}_{\bar{\alpha}\beta \bar{\alpha}_1\beta_1\dots \bar{\alpha}_n\beta_n} =}{}
-\overline{\Omega}_{\bar{\imath}_1 \dots \bar{\imath}_n}\partial^{\bar{\imath}_1}\big(\bar{s}_{\bar{\alpha}_1}s_{\beta_1} {\rm e}^{-\kappa K}\big)\cdots\partial^{\bar{\imath}_n}\big(\bar{s}_{\bar{\alpha}_n}s_{\beta_n} {\rm e}^{-\kappa K}\big) \Big),
\end{gather*}
with $\partial^{i}=g^{i\bar{\jmath}}\bar{\partial}_{\bar{\jmath}}$ and $\partial^{\bar{\imath}}=g^{\bar{\imath}j}\partial_{j}$.
If ${\rm Herm}(\CH_\kappa)$ is the space of hermitian matrices in $\CH_\kappa$ and ${\rm Comm}(\oplus^{n}{\rm Herm}(\CH_\kappa))$ is
the space of $n$ mutually commuting tuples of hermitian matrices in $\CH_{\kappa}$, we can write the map as ${\cal I}\colon
{\rm Comm}(\oplus^{n}{\rm Herm}(\CH_\kappa))\to {\rm Herm}(\CH_\kappa)$. Therefore, one can use the kernel of
${\cal I}$ to approximate special Lagrangian submanifolds as the level sets of the $n$ functions ``$\langle \tilde{\Omega}_{x,\kappa}\vert{\rm ker}({\cal I})\vert \tilde{\Omega}_{x,\kappa}\rangle$''.

Also, one can generalize this quantum system by coupling the particle to a rank $r$ holomorphic vector bundle $V\to M$. We will not give many details of this generalization here, although we will say a few words. For instance, the system can be interpreted as a particle endowed with certain $U(r)$-charge. The associated quantum Hilbert space is $H^{0}(M,V\otimes\CL^{\otimes\kappa})$. One can also def\/ine an analogous set of coherent states and an associated distortion function.
In the semiclassical limit, requiring the generalized distortion function to be constant as a function of $M$ gives rise to generalized balanced metrics, and therefore, to hermite-Yang--Mills metrics on $V\to M$ when $\kappa^{-1}=0$~\cite{DKLR:i}.

Finally, as a technical comment, the \emph{balanced metric} equations \eqref{norm} and \eqref{FS} can be explicitly solved for f\/inite $\kappa$, and its solutions used to approximate Ricci-f\/lat metrics and hermitian Yang--Mills connections. A method to solve them involves the concepts of T-map and algebraic Monte-Carlo integration \cite{Don:iii, DKLR:ii}, which can be applied whenever one has enough analytical control on the Kodaira's embeddings $M\hookrightarrow \IP H^{0}(M,\CL^{\otimes\kappa})$. We leave the problem of developing technical methods for constructing
special Lagrangian submanifolds and other geometric objects for future work. In the following sections we will show how the concept of \emph{balanced metric} appears naturally in other frameworks for quantization (Berezin's star product and path integral quantization), and thus gives rise to K\"ahler--Einstein metrics in the classical limit.

\section{Berezin's star product}\label{section3}

Instead of quantizing the space of observables by introducing a Hilbert space of states, $\CH_{\kappa}$, and its corresponding space of quantum observables (i.e., the hermitian matrices), one can understand quantization as a noncommutative deformation of the geometry of $M$. In the deformation quantization approach to noncommutative geometry, the ordinary algebra of functions $C^{\infty}(M)$ is replaced by the noncommutative $\star$ algebra $C^{\infty}(M)[[\kappa^{-1}]]$, which ref\/lects the operator algebra of hermitian operators on $\CH_{\kappa}$. The $\star$ product of two elements in $C^{\infty}(M)[[\kappa^{-1}]]$ is def\/ined through formal series expansions in powers of $\kappa^{-1}$, such that,
\[
[f, g]:= f\star g - g\star f = i\kappa^{-1}\{f, g\}_{PB} + O\big(\kappa^{-2}\big).
\]
The explicit form of the algebra is not unique \cite{Karabegov:1995wk}, in the same way that the quantization of a classical system is not unique. Here, we will f\/irst explore the Reshetikhin--Takhtajan star product in K\"ahler geometry~\cite{nikolai}.

To describe this algebra, we f\/irst introduce the diagonal of the Bergman kernel and the Calabi's diastatic function. Using the notation introduced above, the diagonal of the Bergman kernel can be written as
\[
e(z,\bar{z}) = \sum_{\alpha} \bar{s}_{\bar{\alpha}}(\bar{z})s_{\alpha}(z) \exp(-\kappa K(z,\bar{z})),
\]
which coincides with the distortion function def\/ined in \eqref{distortion}. The Calabi function is simply
\[
\phi(z,\bar{z};v,\bar{v}) = K(z,\bar{v}) + K(v,\bar{z}) - K(z,\bar{z}) - K(v,\bar{v}).
\]
Note that $e(z,\bar{z})$ and $\phi(z,\bar{z};v,\bar{v})$ are invariant under K\"ahler transformations
$K\to K + f +\bar{f}$. Using the Berezin's formula, one can def\/ine a non-normalized product given by
\[
(f\bullet g)(z,\bar{z}) := \int_{M} f(z,\bar{v})g(v,\bar{z}) \exp \left(-\kappa \phi(z,\bar{z};v,\bar{v})\right) {\omega^n \over n!},
\]
which can be used to introduce the normalized product
\[
(f\star g)(z,\bar{z}) = \int_{M} f(z,\bar{v}) g(v,\bar{z}){e(z,\bar{v})e(v,\bar{z})\over e(z,\bar{z})}
\exp \left(-\kappa \phi(z,\bar{z};v,\bar{v})\right) {\omega^n \over n!}.
\]
The Calabi's diastatic function is def\/ined in some neighborhood of the diagonal $M\times M$, and the point $v=z$ is a critical point of the Calabi function considered as a function of $v$ and $\bar{v}$;  the Laplace expansion of $e^{-\kappa\phi}$ at the critical point $v=z$ yields a formal power series in $\kappa^{-1}$. As it is shown in \cite{nikolai}, one can determine naturally the $\bullet$ product as a combinatoric expansion in powers of $\kappa^{-1}$, derived from the Laplace expansion of the diastatic function. Therefore, one has to compute the $\bullet$ product in order to determine the normalized $\star$ product. The unit element of their noncommutative deformation $C^{\infty}(M)[[\hbar]]$ given by the $\bullet$ product is the diagonal of the Bergman kernel $e(z,\bar{z})$.

Therefore, as a corollary, if the unit element of the Reshetikhin and Takhtajan algebra is constant, i.e.\ the corresponding Bergman kernel has a constant diagonal, the metric on $M$ is balanced. This shows how balanced metrics become natural objects in deformation quantization. One can study this phenomenon further in the path integral quantization formalism.

\section{Completely degenerated quantum systems}\label{section4}

In this section we compute the quantum vacuum energy density associated to a constant Hamiltonian function on $M$, in the semiclassical limit, $\hbar=\kappa^{-1}\to 0$. In the geometric quantization framework, the quantum Hamiltonian associated with the classical Hamiltonian function ${\bf 1}\colon M\to 1$, is the identity operator ${\rm Id}$; if we identify the semiclassical vacuum states with the coherent states, the semiclassical vacuum energy density will be proportional to $\langle \tilde\Omega_x \vert \tilde\Omega_x \rangle$. However, in the path integral approach, the Hamiltonian is set to be zero, and the choice of vacuum is not necessarily the same as the identif\/ication ``coherent state'' = ``vacuum state.''

By subtracting the classical energy density to $\langle \tilde\Omega_x \vert \tilde\Omega_x \rangle$, we will compare the path integral approach and the geometric quantization approach, and f\/ind that the leading term in $\kappa^{-1}$
is the same, though the f\/irst sub-leading correction is not. This means that requiring the semiclassical vacuum energies (in both quantization frameworks) to be constant, yields metrics on $M$ that become K\"ahler--Einstein in the classical limit. The fact that the sub-leading corrections are dif\/ferent only af\/fects higher corrections to the aforementioned metrics when $\kappa^{-1}$, though small, it is not zero.

\subsection{Vacuum energy in geometric quantization}

The system is completely degenerated when the Hamiltonian function is constant; each point in the phase space is a \emph{classical vacuum state} and the quantum Hilbert space becomes the space of \emph{quantum vacua}.
On the {\it geometric quantization} side, one identif\/ies the quantum Hilbert space~$\CH_\kappa$ with the space of holomorphic sections $H^{0}(M, \CL^{\otimes k})$. The natural candidate to be the semiclassical quantum vacuum state peaked at $x$ is the coherent state introduced by Rawnsley~\cite{Rawnsley:1976}, and denoted by $\vert \tilde \Omega_x \rangle$. We construct $\vert \tilde \Omega_x \rangle$ as follows: f\/irst, we choose an orthonormal basis of holomorphic sections, $\{ s_\alpha \in H^{0}(\CL^{\otimes k}) \}_1^{N(\kappa)}$ with
\[
\langle s_{\alpha}\vert s_{\beta}\rangle = \delta_{\bar{\alpha}\beta} = \int_{M} \bar{s}_{\bar{\alpha}}(\bar{x}) s_{\beta}(x) {\rm e}^{ -\kappa K(x,\bar{x})}
{\omega^n(x,\bar{x}) \over n!},
\]
where $s_{\alpha}(x){\rm e}^{ -\kappa K(x,\bar{x})/2}$ is the complex number associated with the evaluation of the holomorphic section $s_\alpha$ at $x\in U$, and def\/ined in the trivialization $\CL^{\otimes\kappa}\vert_U\simeq \IC\times U$. Second and lastly, one can def\/ine the \emph{coherent state} peaked at $x\in M$ as the ray in $\IP\CH_\kappa$ generated by
\[
\vert \tilde \Omega_x \rangle := \sum_{\alpha=1}^{N(\kappa)} s_{\alpha}(x){\rm exp}\left( -\kappa K(\bar{x},x)/2\right)\vert s_{\alpha} \rangle \in \CH_\kappa,\quad x\in U\subset M,
\]
and one can easily show how such a ray is independent of the choice of trivialization of the line bundle.

The set of coherent states $\{ \vert \tilde \Omega_x \rangle \}_{x\in M}$ is a supercomplete system of vectors in $\CH_\kappa$, parametrized by the points of $M$. It also def\/ines an embedding of $M$ into $\IP\CH_\kappa$, and
implies the Parseval identity~\eqref{parseval}.
This allows the def\/inition of an embedding of the space of quantum observables in~$\CH_\kappa$ (i.e., the self-adjoint matrices in ${\rm Herm}(\CH_\kappa)\subset\CH_\kappa^{\ast}\otimes\CH_\kappa$) into the space of classical observables $C^{\infty}(M)$, according to the formula
\[
\hat{f}\mapsto \langle \tilde \Omega_{\circ} \vert \hat{f} \vert \tilde \Omega_{\circ} \rangle = f \in C^{\infty}(M), \qquad \hat{f}=\hat{f}^\dagger,
\]
where $\circ$ denotes the pre-image of $f$ in $M$. The function $f$ is called a \emph{covariant} symbol of the matrix
$\hat{f}$. A function $\check{f} \in C^{\infty}(M)$ such that the matrix $\hat{f}$ is representable as
\[
\hat{f} = \int_{M} \vert \tilde \Omega_x \rangle\otimes \langle \tilde \Omega_x \vert {\check{f} \omega^n(x,\bar{x}) \over n!},
\]
is called \emph{contravariant} symbol of the matrix $\hat{f}$. The map $\check{f}\mapsto \hat{f}$ is also known as \emph{Toeplitz map}, $T\colon  C^{\infty}(M) \to {\rm Herm}(\CH_\kappa) \subset \CH_\kappa^{\ast}\otimes\CH_\kappa$.

It is important to stress that the identif\/ication between the \emph{quantum vacuum state}  $\vert  \Omega_x\rangle$ localized at $x\in M$ and the \emph{coherent state} $\vert \tilde \Omega_x\rangle$ is, in somehow, made arbitrarily.
Such an identif\/ication is motivated by the fact that the coherent state is peaked at $x$ and localized within a neighborhood $V_x \subset M$ with minimal quantum uncertainty. More precisely, the
coherent state $\vert \tilde \Omega_{x,\kappa}\rangle$ satisf\/ies
\begin{gather}
{1\over \langle  \tilde \Omega_{x,\kappa} \vert \tilde \Omega_{x,\kappa}\rangle}\int_{V_x\subset M} \langle  \tilde \Omega_{x,\kappa} \vert  \tilde \Omega_{y,\kappa}\rangle \langle  \tilde \Omega_{y,\kappa} \vert  \tilde \Omega_{x,\kappa}\rangle  {\omega^n(y,\bar{y}) \over n!} \sim 1,
\label{uncertainty}
\end{gather}
with $\int_{V_x} \omega^n / \int_M\omega^n \sim {1\over N}$. Such an identif\/ication
is correct as a f\/irst approximation in $\kappa^{-1}$, although the $O(\kappa^{-2})$ terms are not universal and depend on the choice of vacuum state $\vert \Omega_{x,\kappa} \rangle$. The $O(\kappa^{-2})$ corrections are important at the time of computing correlation functions of the type
\begin{gather}
\langle \Omega_{x,\kappa} \vert \hat{f}_1 \hat{f}_2\cdots \hat{f}_m \vert \Omega_{x,\kappa} \rangle
\label{correlator}
\end{gather}
in the limit $\kappa\!\to\!\infty$, as power series in $\kappa^{-1}$ of the covariant symbols  $f_1, f_2,\ldots, f_m \!\in\! C^{\infty}(M)[[\kappa^{-1}]]$. For instance, every identif\/ication of the vacuum state with peaked states that obey equation~\eqref{uncertainty}, gives rise to the same semiclassical limit
\[
\langle \Omega_{x,\kappa} \vert [\hat{f}, \hat{g} ] \vert \Omega_{x,\kappa} \rangle =
i\kappa^{-1}\omega^{i\bar{\jmath}}\left( \partial_i f(x,\bar{x}) \bar{\partial}_{\bar{\jmath}} g(x,\bar{x}) - \partial_i
g(x,\bar{x}) \bar{\partial}_{\bar{\jmath}}f(x,\bar{x}) \right) + O\big(\kappa^{-2}\big),
\]
with $f(x,\bar{x}) := \langle \Omega_{x,\kappa}\vert \hat{f} \vert \Omega_{x,\kappa} \rangle $
and $g(x,\bar{x}) :=\langle \Omega_{x,\kappa}\vert \hat{g} \vert \Omega_{x,\kappa} \rangle$,~\cite{Bordemann:i}. However, the higher corrections $O(\kappa^{-2})$ will depend on the choice of vacuum state.

To compute the semiclassical limit of the correlators \eqref{correlator} beyond $O(\kappa^{-1})$ is a dif\/f\/icult task which involves hard analysis; see~\cite{Lu:2000} for the most recent results.
For simplicity, we study only the vacuum expectation value of the identity operator using the na\"ive vacuum state $\vert \tilde \Omega_{x,\kappa} \rangle$. Z.~Lu computed the lower order terms in powers of $\kappa^{-1}$ of
the squared norm of the na\"ive vacuum state,~\cite{Lu:2000},
\begin{gather*}
\langle \tilde \Omega_{x,\kappa} \vert \tilde \Omega_{x,\kappa} \rangle = 1 + {1\over 2\kappa}R + {1\over 3\kappa^2}\left(\Delta R +
{1\over 8}\left(\vert \mathrm{Riemann} \vert^2 -4\vert \mathrm{Ricci} \vert^2 + 3 R^2\right)\right) + O\big(\kappa^{-3}\big).
\end{gather*}
Computing the asymptotic series of the vacuum energy on the path integral side involves perturbative expansions of Feynman vacuum diagrams. As the classical energy density is set to be zero in the path integral formalism, one should compare the path integral result to the ``renormalized'' geometric quantized vacuum energy
\begin{gather}
E_0(x)=\langle \tilde \Omega_{x,\kappa} \vert \tilde \Omega_{x,\kappa} \rangle-1 \nonumber\\
\phantom{E_0(x)}{}
= {1\over 2\kappa}R + {1\over 3\kappa^2}\left(\Delta R +
{1\over 8}\left(\vert \mathrm{Riemann} \vert^2 -4\vert \mathrm{Ricci} \vert^2 + 3 R^2\right)\right) + O\big(\kappa^{-3}\big),\label{expansion2}
\end{gather}
which is zero at $\kappa=\infty$.

\subsection{Path integral derivation of the vacuum energy}

On the \emph{path integral quantization} side, Cattaneo and Felder \cite{Cattaneo:1999} give a prescription for computing correlation functions for quantized observables $f, g\in C^{\infty}(M)[[\kappa^{-1}]]$, by evaluating path integrals perturbatively as formal expansions in powers of
$\kappa^{-1}$.
In such perturbative expansion one considers perturbations around the constant map, i.e., the solution of the equations of motion or \emph{classical vacuum state} $\Phi_0 \colon \IR\mapsto x\in M$. For simplicity, we choose a local coordinate chart around~$x$, $\{\phi_i \}_{0< i \leq n}$ given by the K\"ahler--Riemann normal coordinates~\cite{AlvarezGaume:1981}.
For a vanishing Hamiltonian, the phase-space action associated with perturbations $\Phi$ around the classical va\-cuum state $\Phi_0$ is the line integral of the $U(1)$-connection on the prequantum bundle $\CL^{\otimes k}$, along the path $\Phi$ in $M$
\begin{gather}
S[\Phi] = i\int_{-\infty}^{\infty}dt \bar{\partial}_{\bar{\jmath}}K(\phi,\bar{\phi}){d\bar{\phi}^{\bar{\jmath}}\over dt},
\label{action}
\end{gather}
with $\Phi \in {\rm Maps}(\IR\to M)$ and\footnote{We denote by ${\rm Maps}(\IR\to M)$ the functional space of maps of the real line $\IR$ to the phase space $M$.} $\Phi(\pm\infty)=x$, implying $\phi(\pm\infty)=0$ in the local K\"ahler--Riemann normal coordinate system around $x$. The functional integration of f\/luctuations around the classical vacuum $x\in M$ def\/ines a semiclassical quantum
vacuum state that we denote as $\vert \Omega_{x,\kappa}\rangle$, although we don't know how to describe it as an element of the Hilbert space, $\CH_{\kappa}$.

The prescription for computing the correlation functions that appear in deformation quantization \cite{Cattaneo:1999, Kontsevich:1997} is given by the path integral on the phase-space variables
\begin{gather}
\langle \Omega_x \vert \hat{f} \hat{g} \vert\Omega_x \rangle := f\star g(x,\bar{x}) =
\int_{{\rm Maps}(\IR\to M)\vert \Phi(\pm\infty)=x} d\Phi\, f(\Phi(1))g(\Phi(0))\exp\left( i\kappa S[\Phi] \right),
\label{starprod}
\end{gather}
with $f, g \in C^{\infty}(M)[[\kappa^{-1}]]$.
In Riemann normal coordinates we can write the action~\eqref{action} as follows
\begin{gather*}
i\kappa S[\Phi]=-\kappa\int_{\IR} dt \bar{\partial}_{\bar{\jmath}}K(\phi,\bar{\phi}){d\bar{\phi}^{\bar{\jmath}}\over dt} =
-\kappa\int_{\IR} dt\Bigg(
g_{i\bar{\jmath}}(x,\bar{x})\phi^i\pdbar^{\bar{\jmath}} + {1\over 2} R_{i\bar{\jmath}k\bar{l}}\phi^i\bar{\phi}^{\jbar}\phi^k\pdbar^{\bar{l}}  \nonumber\\
\phantom{i\kappa S[\Phi]=}{}
+ {1\over 6} D_{m}R_{i\bar{\jmath}k\bar{l}}\phi^m\phi^i\bar{\phi}^{\jbar}\phi^k\pdbar^{\bar{l}} + {1\over 6} \bar{D}_{\bar{m}}R_{i\bar{\jmath}k\bar{l}}\bar{\phi}^{\bar{m}}\phi^i\bar{\phi}^{\jbar}\phi^k\pdbar^{\bar{l}}
+ {1\over 12} D_{n} D_{m} R_{i\bar{\jmath}k\bar{l}}\phi^n \phi^m\phi^i\bar{\phi}^{\jbar}\phi^k\pdbar^{\bar{l}} \nonumber\\
\phantom{i\kappa S[\Phi]=}{}+ {1\over 12} \bar{D}_{\bar{n}} \bar{D}_{\bar{m}} R_{i\bar{\jmath}k\bar{l}}\bar{\phi}^{\bar{n}} \bar{\phi}^{\bar{m}}\phi^i\bar{\phi}^{\jbar}\phi^k\pdbar^{\bar{l}}
+ {1\over 12}  \bar{D}_{(\bar{n}} D_{m} R_{i\bar{\jmath}k\bar{l})}\phi^m \bar{\phi}^{\bar{n}}\phi^i\bar{\phi}^{\jbar}\phi^k\pdbar^{\bar{l}} \nonumber\\
\phantom{i\kappa S[\Phi]=}{}+ {1\over 4} g^{o\bar{r}} R_{o(\bar{\jmath}m\bar{l}}  R_{i\bar{n}k)\bar{r}}  \phi^m \bar{\phi}^{\bar{n}}\phi^i\bar{\phi}^{\jbar}\phi^k\pdbar^{\bar{l}}
+ O\big(\phi^7\big)\cdots \Bigg) ,
\end{gather*}
where $\dot{\phi}=d\phi/dt$, and the parentheses enclosing indices indicate the completely symmetric part of such indices; we sum over repeated indices.  Still, the measure $d\Phi$ in the functional integration~\eqref{starprod} also depends of the phase-space coordinate f\/ield $\phi$. Hence, in this choice of coordinates
\begin{gather}
{\rm e}^{-iE_{0}^{\prime}(x)\delta(0)} =\! \int_{{\rm Maps}(\IR\to M) \vert \Phi(\pm\infty)=x}\! \prod_{i=1}^n d\phi^i \bar{d}\bar{\phi}^{\bar{i}}
\det \omega\left(\phi,\bar{\phi}\right) \exp\!\left(\! -\kappa\!\int_{\IR} dt \bar{\partial}_{\bar{\jmath}}K(\phi,\bar{\phi}){d\bar{\phi}^{\bar{\jmath}}\over dt} \right),\!\!\!\!
\label{dddd}
\end{gather}
where $\det\omega$ is the determinant of the K\"ahler form $\omega_{i\bar{\jmath}}$, and $E_{0}^{\prime}(x)$ is the quantum vacuum energy density, depending on the choice of semiclassical vacuum labeled as $x\in M$. Therefore, as
$\omega$ depends on the integration variables, we can introduce an anti-commuting auxiliary f\/ield $\lambda$
to write the path integral using a standard gaussian measure
\begin{gather*}
{\rm e}^{-iE_{0}^{\prime}(x)\delta(0)} = \int \prod_{i=1}^n d\phi^i \bar{d}\bar{\phi}^{\bar{i}} d\lambda^i \bar{d}\bar{\lambda}^{\bar{i}} \exp\left( -\kappa\int_{\IR} dt \bar{\partial}_{\bar{\jmath}}K(\phi,\bar{\phi}){d\bar{\phi}^{\bar{\jmath}}\over dt} + \int_{\IR} dt\, \omega( \phi,\bar{\phi})_{i\bar{\jmath}}\lambda^i \bar{\lambda}^{\bar{\jmath}}
\right),
\end{gather*}
where the functional integral of the auxiliary f\/ield $\lambda$ obeys the rules of the Grassmann integration. One can expand the action for the auxiliary f\/ield, in powers of the f\/ield $\phi$, to f\/ind out the interactions between auxiliary f\/ield and phase-space coordinate f\/ield,
\begin{gather*}
 \int_{\IR} dt\, \omega( \phi,\bar{\phi})_{i\bar{\jmath}}\lambda^i \bar{\lambda}^{\bar{\jmath}} =
i\int_{\IR} dt  \Bigg( g_{i\bar{\jmath}}\lambda^i \bar{\lambda}^{\bar{\jmath}} + R_{i\bar{\jmath}k\bar{l}}\lambda^i \bar{\lambda}^{\bar{\jmath}}\phi^k\pdbar^{\bar{l}}
+
{1\over 12}  \bar{D}_{(\bar{n}} D_{m} R_{i\bar{\jmath}k\bar{l})}\lambda^i \bar{\lambda}^{\bar{\jmath}} \phi^m \bar{\phi}^{\bar{n}}\phi^k\pdbar^{\bar{l}}
 \nonumber\\
\hphantom{\int_{\IR} dt\, \omega( \phi,\bar{\phi})_{i\bar{\jmath}}\lambda^i \bar{\lambda}^{\bar{\jmath}} =}{}
+ {1\over 4} g^{o\bar{r}} R_{o(\bar{\jmath}m\bar{l}}  R_{i\bar{n}k)\bar{r}}  \lambda^i \bar{\lambda}^{\bar{\jmath}}\phi^m \bar{\phi}^{\bar{n}}\phi^k\pdbar^{\bar{l}} + \cdots \Bigg),
\end{gather*}
where the coef\/f\/icients $g_{i\bar{\jmath}}$, $R_{i\bar{\jmath}k\bar{l}}$, etc., are evaluated at $x\in M$.

Therefore, we can evaluate $E_{0}^{\prime}(x)$ perturbatively as an expansion of the path integral in powers of $\kappa^{-1}$,
in the limit $\kappa\to\infty$. If we write the Fourier transform of the f\/ield to the momentum variables as
\begin{gather*}
\hat{\phi}(p) = {1\over 2\pi}\int_{\IR} dt\,\exp(ipt)\phi(t), \qquad \phi(\pm\infty)=0,
\\
\hat{\lambda}(p) = {1\over 2\pi}\int_{\IR} dt\,\exp(ipt)\lambda(t), \qquad \lambda(\pm\infty)=0,
\end{gather*}
the propagators in the momentum space are
\begin{figure}[h!]
\centerline{\includegraphics[width=3.3cm]{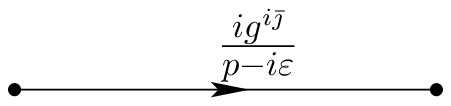} \qquad\qquad
    \includegraphics[width=3.3cm]{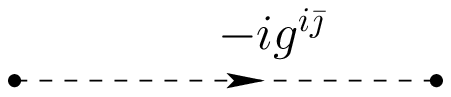}}
\caption{Propagators for the phase-space coordinate f\/ield (left), and the auxiliary f\/ield (right).}
\label{i}
\end{figure}

\begin{figure}[t]
\centerline{    \includegraphics[height=3.3cm]{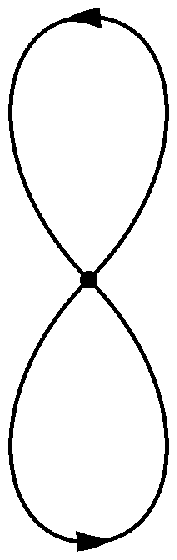} \qquad\qquad
    \includegraphics[height=3.3cm]{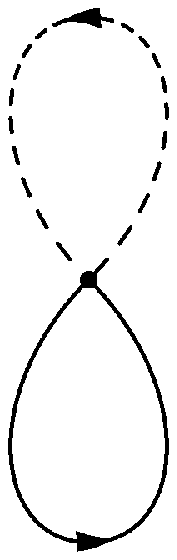}}
\caption{Two-loop vacuum diagrams, (2.i) left and (2.ii) right.}
\label{iv}
\end{figure}

\begin{figure}[t]
\centerline{\includegraphics[width=4.3cm]{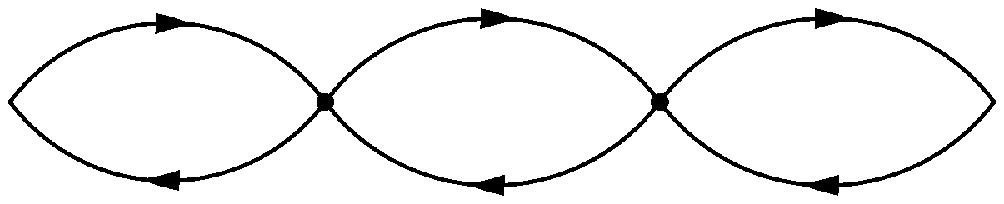}}
\caption{Three-loop vacuum diagram (3.i).}
\label{v}
\end{figure}

\begin{figure}[t]
\centerline{\includegraphics[width=3.3cm]{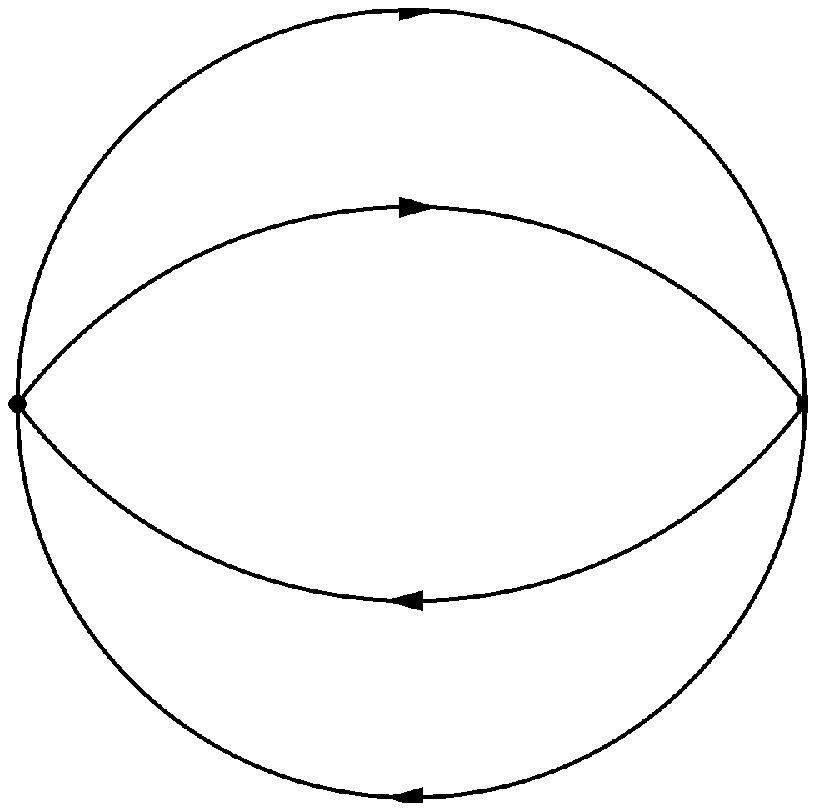}}
\caption{Three-loop vacuum diagram (3.ii).}
\label{vi}
\end{figure}

\begin{figure}[t]
\centerline{\includegraphics[width=3.3cm]{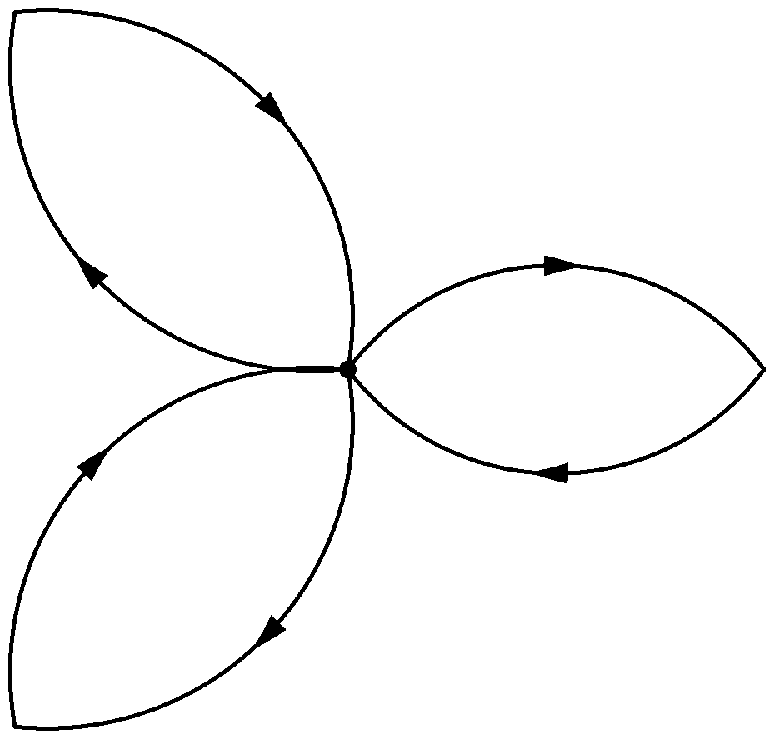}}
\caption{Three-loop vacuum diagram (3.iii).}
\label{vii}
\end{figure}

\begin{figure}[t]
\centerline{\includegraphics[width=4.3cm]{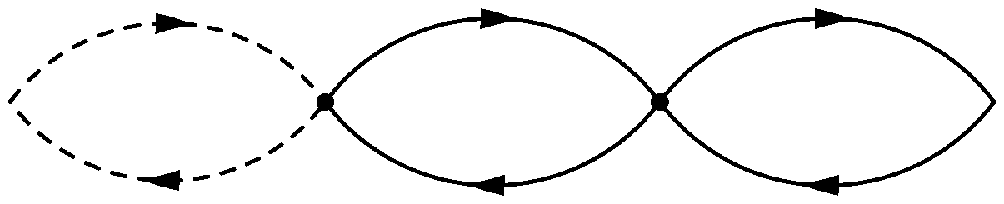} \qquad\qquad
    \includegraphics[width=4.3cm]{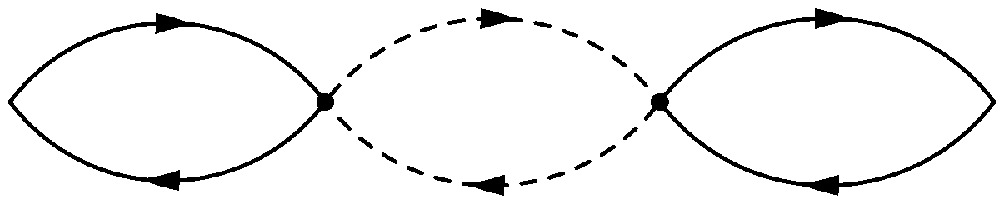}}
\caption{Three-loop vacuum diagrams with auxiliary f\/ield, (3.iv) left and (3.v) right.}
\label{ix}
\end{figure}

\begin{figure}[t]
\centerline{\includegraphics[width=3.3cm]{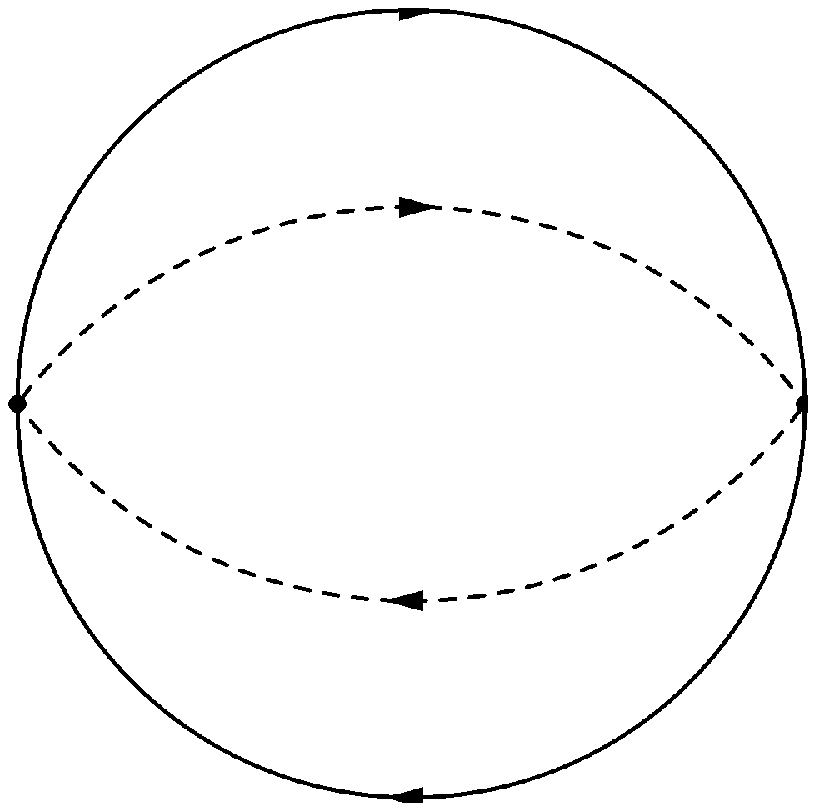}}
\caption{Three-loop vacuum diagram with auxiliary f\/ield (3.vi).}
\label{viii}
\end{figure}

\begin{figure}[t!]
\centerline{\includegraphics[width=3.3cm]{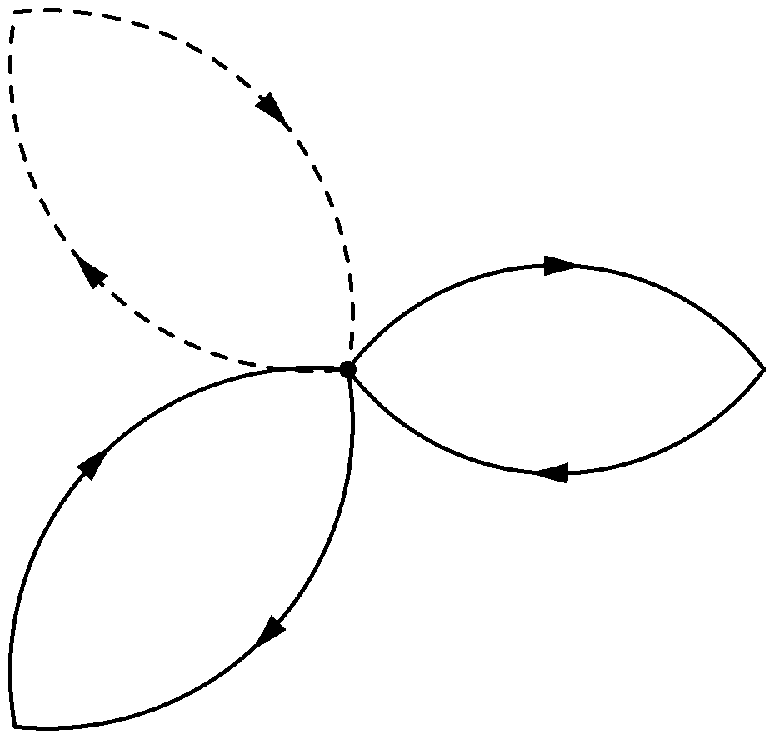}}
\caption{Three-loop vacuum diagram with auxiliary f\/ield (3.vii).}
\label{x}
\end{figure}

In order to compute \eqref{dddd}, we perform a perturbative expansion in powers of $\kappa^{-1}=\hbar$. We can compute the vacuum energy $E_{0}^{\prime}(x)$, by simply summing the connected vacuum (or bubble) diagrams, and dividing by the total length of $\IR$. Thus, a vacuum diagram with $L$ \emph{loops} contributes to $E_{0}^{\prime}(x)$ a term proportional to $\kappa^{1-L}=\kappa^{V-P}$, with $V$ the \emph{number of vertices} and $P$ the \emph{number of propagators}. Therefore, to determine $E_{0}^{\prime}(x)$ up to order $\kappa^{-2}$, we have to sum the connected diagrams depicted in Figs.~\ref{iv}--\ref{x}.

\begin{figure}[h!]
\centerline{\includegraphics[height=2.5cm]{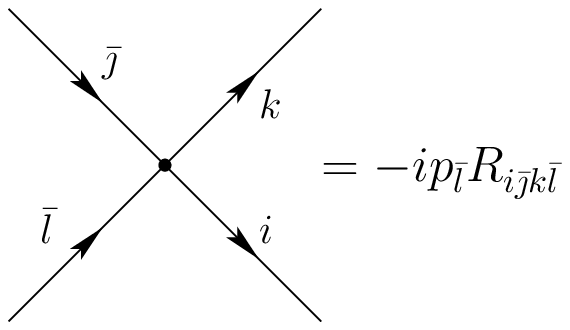} \qquad\qquad
    \includegraphics[height=2.5cm]{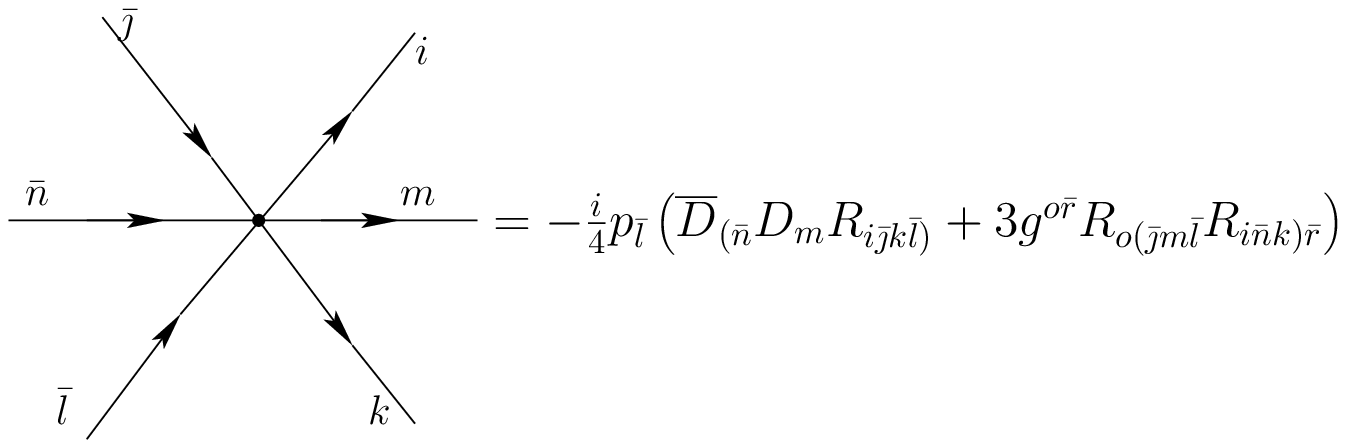}}
\caption{Interaction vertices for the phase-space f\/ield. $p_{\bar{l}}$ denotes the momentum carried by the particle which propagates along the $\bar{l}$-leg.}
\label{ii}
\end{figure}

\begin{figure}[h!]
\centerline{\includegraphics[height=2.5cm]{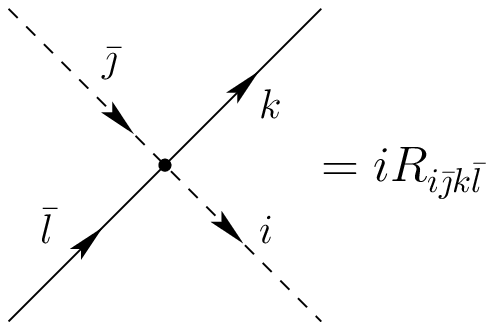} \qquad\qquad
    \includegraphics[height=2.5cm]{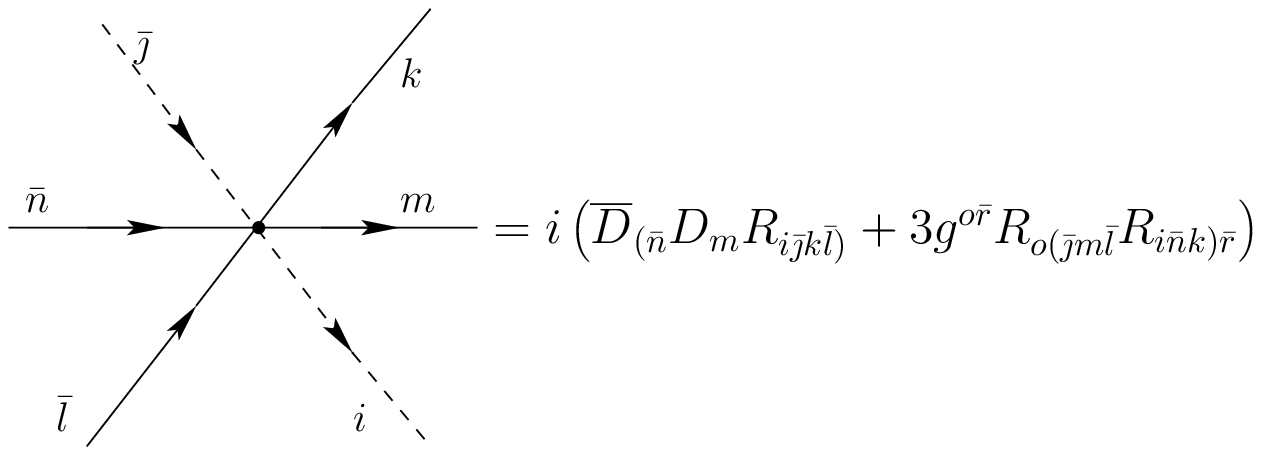}}
\caption{Interaction vertices for the auxiliary f\/ield with the phase-space f\/ield.}
\label{iii}
\end{figure}

As we want only to evaluate diagrams up to order $\kappa^{-2}$, we only need to consider a few interactions; by drawing the diagrams, one realizes that the only vertices that appear are the ones depicted in Figs.~\ref{ii} and~\ref{iii}. The asymptotic expansion of $E_0^{\prime}(x)$ in powers of $\kappa^{-1}$ can be expressed as
\begin{gather}
-iE_0^{\prime}(x)\delta(0) = \sum_{\Gamma \in \mathfrak{G}_2} {1 \over \kappa \vert {\rm Aut}(\Gamma) \vert}D_\Gamma (x) +
\sum_{\Gamma \in \mathfrak{G}_3} {1 \over \kappa^2 \vert {\rm Aut}(\Gamma) \vert}D_\Gamma (x) + O\big(\kappa^{-3}\big),
\label{asymptoticexp}
\end{gather}
where $ \mathfrak{G}_L$ is the set of bubble diagrams with $L$ loops,  ${\rm Aut}(\Gamma)$ is the subgroup of the
group of automorphisms of $\Gamma$ that maps vertices to vertices of the same type and oriented propagators to oriented propagators of the same type (which start and end at the same vertices), $\vert {\rm Aut}(\Gamma) \vert = \# {\rm Aut}(\Gamma)$ is also known as \emph{symmetry factor}, and $D_\Gamma(x)$ is the evaluation of the Feynman diagram.

The evaluation of each diagram $D_\Gamma(x)$ follows from the Feynman rules in momentum space: to each line we associate its corresponding propagator (Fig.~\ref{i}), to each vertex we associate its corresponding numerical factor (Figs.~\ref{ii} and~\ref{iii}), we impose momentum conservation at each vertex and integrate
over each undetermined momentum $\int_{\IR}{dp\over 2\pi}$. There are two types of integrals that appear in the evaluation of bubble diagrams
\[
\lim_{\varepsilon\to 0^{+}} {1\over 2\pi}\int_{\IR} {idp\over p-i\varepsilon} = {1 \over 2} \qquad {\rm and}\qquad \delta(0) = {1\over 2\pi} \int_{\IR} dp.
\]
Each vacuum diagram is proportional to the Dirac delta $\delta(0)$, or the ``total length'' of $\IR$, because the calculation in the momentum space yields the total vacuum energy in $\IR$. As we are just interested in the vacuum energy \emph{density}, we will divide out by inf\/inite total length of the ($0+1$)-spacetime, $\IR$. Thus, in order to determine $E_{0}^{\prime}(x)$ up to three loops, we use equation~\eqref{asymptoticexp}. The evaluation of the two- and three-loop diagrams gives rise to the following: for two-loops,
\begin{itemize}\itemsep=0pt
\item (2.i) in Fig.~\ref{iv}
\[
D_{\rm (2.i)}(x)=-i R_{i\bar{\jmath}k\bar{l}}g^{i\bar{\jmath}}g^{k\bar{l}}{1\over (2\pi)^2}\int_{\IR}dp_1{ip_1\over p_1-i\varepsilon}\int_{\IR}dp_2{i\over p_2-i\varepsilon} = \delta(0){1\over 2}R.
\]
\item (2.ii) in Fig.~\ref{iv}
\[
D_{\rm (2.ii)}(x)=i R_{i\bar{\jmath}k\bar{l}}{1\over (2\pi)^2}\int_{\IR}dp_1 (-i)g^{k\bar{l}}\int_{\IR}dp_2{ig^{i\bar{\jmath}}\over p_2-i\varepsilon} = \delta(0) \frac 12 R.
\]
\end{itemize}

For three-loop diagrams,
\begin{itemize}
\item (3.i) in Fig.~\ref{v}
\begin{gather*}
D_{\rm (3.i)}=(-i)^2 R_{i\bar{\jmath}k\bar{l}}g^{i\bar{j}}R_{m\bar{n}o\bar{p}}g^{m\bar{n}}g^{k\bar{p}}g^{o\bar{l}} \int_{\IR^3}{dp_1 dp_2 dp_3\over (2\pi)^3}{i\over p_1-i\varepsilon}\left( {i\over p_2-i\varepsilon} \right)^2 p_2 {ip_3\over p_3-i\varepsilon} \\
\phantom{D_{\rm (3.i)}}{} = \delta(0){1\over 4}\vert {\rm Ricci}\vert ^2.
\end{gather*}

\item (3.ii) in Fig.~\ref{vi}
\begin{gather*}
D_{\rm (3.ii)}=(-i)^2 R_{i\bar{\jmath}k\bar{l}}R^{i\bar{\jmath}k\bar{l}}\int_{\IR^4}{dp_1 dp_2 dp_3 dp_4\over (2\pi)^4}\delta(p_1+p_3-p_2-p_4)p_1 p_2\\
\phantom{D_{\rm (3.ii)}=}{}\times
{i\over p_1-i\varepsilon}{i\over p_2-i\varepsilon}  {i\over p_3-i\varepsilon} {i\over p_4-i\varepsilon} \\
\phantom{D_{\rm (3.ii)}}{}
= -\vert {\rm Riemann} \vert^2\int_{\IR^3}{dp_1 dp_2 dp_3\over (2\pi)^3} {1\over (p_3-i\varepsilon)[(p_1+p_3-p_2)-i\varepsilon]} \\
\phantom{D_{\rm (3.ii)}}{}
=   -\vert {\rm Riemann} \vert^2 \int_{\IR^3} {dl_1 dl_2 dl_3\over (2\pi)^3}{i\over (l_1-i\varepsilon)(l_2-i\varepsilon)} = \delta(0){1\over 4}\vert {\rm Riemann} \vert^2,
\end{gather*}
where $l_1 = p_1+p_2-p_3$, $l_2=p_3$ and
$l_3=p_2$.
\end{itemize}

\begin{itemize}
\item (3.iii) in Fig.~\ref{vii}
 \begin{gather*}
 D_{\rm (3.iii)}={-i\over 4}\left( \Delta R + 3\left( {2\over 3} \vert {\rm Ricci} \vert^2 + {1\over 3} \vert {\rm Riemann} \vert^2\right) \right)\\
 \phantom{D_{\rm (3.iii)}=}{}\times
 \lim_{\varepsilon\to 0^{+}}\int_{\IR^3}{dp_1 dp_2 dp_3\over (2\pi)^3} {i\over p_1-i\varepsilon}  {i\over p_2-i\varepsilon}{ip_3\over p_3}\\
 \phantom{D_{\rm (3.iii)}}{}
 =\delta(0){1\over 16}\left( \Delta R +  2 \vert {\rm Ricci} \vert^2 + \vert {\rm Riemann} \vert^2 \right).
 \end{gather*}
\item (3.iv) in Fig.~\ref{ix}. Similarly to (3.i)
\[
D_{\rm (3.iv)}= \delta(0){1\over 4} \vert {\rm Ricci}\vert^2.
\]
\item (3.v) in Fig.~\ref{ix}
\[
D_{\rm (3.v)}= \delta(0){1\over 4} \vert {\rm Ricci}\vert^2.
\]
\end{itemize}

\begin{itemize}

\item (3.vi) in Fig.~\ref{viii}. Similarly to (3.ii),
\begin{gather*}
D_{\rm (3.vi)}= (iR_{i\bar{m}k\bar{o}})(iR_{n\bar{\jmath}p\bar{l}})\big(-i g^{i\bar{\jmath}}\big) \big(-i g^{k\bar{l}}\big)\int_{\IR^3}{dp_1 dp_2 dp_3\over (2\pi)^3}{ig^{n\bar{m}}\over p_1 - i\varepsilon}{ig^{p\bar{o}}\over p_2 - i\varepsilon}\\
\phantom{D_{\rm (3.vi)}}{}
= \delta(0){1\over 4} \vert {\rm Riemann}\vert^2.
\end{gather*}

\item (3.vii) in Fig.~\ref{x}.
Similarly to (3.iii),
 \begin{gather*}
 D_{\rm (3.vii)} = i\left( \Delta R + 3\left( {2\over 3} \vert {\rm Ricci} \vert^2 + {1\over 3} \vert {\rm Riemann} \vert^2\right) \right)\int_{\IR^3}{dp_1 dp_2 dp_3\over (2\pi)^3} {i\over p_1-i\varepsilon} {i\over p_2-i\varepsilon}  \\
 \phantom{D_{\rm (3.vii)}}{}
   = \delta(0){1\over 4}\left( \Delta R + 2 \vert {\rm Ricci} \vert^2 + \vert {\rm Riemann} \vert^2 \right).
  \end{gather*}
\end{itemize}

Finally, including the \emph{symmetry factors} of each diagram, and summing them as in equation~\eqref{asymptoticexp}, yields the vacuum energy density associated to the semiclassical vacuum state localized at $x\in M$,
\begin{gather}
E_{0}^{\prime}(x) = {1\over 2\kappa}R + {1\over 96\kappa^2}\left( 5\Delta R + 42 \vert {\rm Ricci} \vert^2 + 17 \vert {\rm Riemann} \vert^2 \right) + O\left( {1\over\kappa^{3}} \right).
\label{vacenergy}
\end{gather}
Thus, comparing equation~\eqref{vacenergy} with the equivalent result in geometric quantization~\eqref{expansion2}, yields dif\/ferent vacuum energy densities $E_{0}(x) \neq E_{0}^{\prime}(x)$, despite the fact that the leading terms are identical. The corollary is an interesting one: f\/ixing $\kappa$ and requiring the quantum vacuum energy density to be constant on the quantum moduli space of semiclassical vacua is equivalent to endowing $M$ with a ``generalized balanced metric''. This generalized notion of balanced metric gives rise to the same K\"ahler--Einstein metrics in the classical limit, which shows that the emergence of K\"ahler--Einstein metrics in the classical limit is generic for a broad choice of semiclassical vacuum states.

\section{Conclusion}\label{section5}

We have shown how the K\"ahler--Einstein metrics appear naturally in the classical limit of K\"ahler quantization. In geometric quantization, identifying semiclassical vacuum states with coherent states allows us to def\/ine balanced metrics as those metrics which yield constant semiclassical vacuum energy (for constant classical Hamiltonian). In the Berezin's approach to deformation quantization, the unit element of the noncommutative algebra $C^{\infty}(M)[[\kappa^{-1}]]$ is the constant function, if and only if the metric is balanced. Also in path integral quantization, requiring the semiclassical vacuum energy to be constant yields a metric that is K\"ahler--Einstein in the classical limit.

Strictly speaking, the metrics that appear in path integral quantization are not balanced. This is due to a dif\/ferent choice of vacuum states in the path integral formalism; thus, for each choice of moduli spaces of semiclassical vacua one can def\/ine dif\/ferent \emph{generalized balanced metrics}. It would be interesting to study the properties exhibited by this general class of metrics. For instance, it is especially interesting to understand how introducing quantum corrections to the K\"ahler potential deforms the moduli of semiclassical vacua~\cite{Karabegov:1995wk}.

Another interesting problem would be to understand balanced metrics in vector bundles within the framework of K\"ahler quantization. Also, one could explicitly construct special Lagrangian submanifolds in Calabi--Yau threefolds, and give a \emph{geometric quantization} formulation of the Bressler--Soilbeman conjecture~\cite{Bressler:2002eu} (which conjectures a correspondence of the Fukaya category with a certain category of holonomic modules over the quantized algebra of functions).

A f\/inal motivation for future research comes from the fact that the geometric objects explored in this paper appear in the large volume limit of string theory compactif\/ications.  We have shown how these objects can be explicitly constructed in the semiclassical limit of geometric quantization; one would expect that dif\/ferent areas of string theory, such as Matrix theory, black holes, and Calabi--Yau compactif\/ication theory~\cite{Banks:1996vh, Cornalba:1998zy,  Gaiotto:2004pc, Kachru:1997bi}, where the quantized algebra of functions plays a special role, could be understood better through a deeper study of the ideas explored here.

\subsection*{Acknowledgements}

It is a pleasure to thank T.~Banks, E.~Diaconescu, M.~Douglas, R.~Karp, S.~Klevtsov, and specially the author's advisor G.~Moore, for valuable discussions. We would like to thank as well G.~Moore and G.~Torroba for their comments on the manuscript, and J.~Nannarone for kind encouragement and support. This work was supported by DOE grant DE-FG02-96ER40949.

\pdfbookmark[1]{References}{ref}
\LastPageEnding

\end{document}